\documentclass[a4paper,11pt]{article}
\usepackage{pos}

\usepackage{multirow}
\usepackage{ctable}
\usepackage{booktabs}
\usepackage{array}
\usepackage{tabularx}
\usepackage{xcolor}
\usepackage{pstricks}
\title{Production of $\nu_{\tau}$ neutrinos and $\overline{\nu}_{\tau}$ antineutrinos - elaborate calculation for a fixed target experiment SHiP}
\ShortTitle{Production of $\nu_{\tau}$ neutrinos and $\overline{\nu}_{\tau}$ antineutrinos at SHiP}

\author*[a]{Rafa{\l} Maciu{\l}a}
\affiliation[a]{Institute of Nuclear
Physics, Polish Academy of Sciences,\\ ul. Radzikowskiego 152, PL-31-342 Krak{\'o}w, Poland}

\emailAdd{rafal.maciula@ifj.edu.pl}
\abstract{We discuss cross sections for 
$\nu_{\tau}$ and ${\overline \nu}_{\tau}$ production
from the direct $D_s^{\pm} \to \nu_{\tau}/{\overline \nu}_{\tau}$ and chain $D_s^{\pm} \to \tau^+/\tau^- \to \nu_{\tau}/{\overline \nu}_{\tau}$ decays in $p\!+^{96}\!\mathrm{Mo}$ scattering with proton beam $E_{\mathrm{lab}}$ = 400 GeV \textit{i.e.} at $\sqrt{s}_{NN}$ = 27.4 GeV.
In our calculations we include $D_s^{\pm}$ from charm fragmentation
$c \to D_s^{+}$ and $\bar c \to D_s^-$ as well as those from
subleading fragmentation of strange quarks/antiquarks $s \to D_s^-$ and $\bar s \to D_s^+$.
The different contributions to $D_s^{\pm}$ and
$\nu_{\tau} / {\overline \nu}_{\tau}$ production rates are shown explicitly.
Estimates of a number of observed 
$\nu_{\tau} / \overline{\nu}_{\tau}$ in 
the $\nu_{\tau} / \overline{\nu}_{\tau} +^{208}\!\mathrm{Pb}$ reaction, 
with 2m long target are given.
}

\FullConference{%
  40th International Conference on High Energy physics - ICHEP2020\\
  July 28 - August 6, 2020\\
  Prague, Czech Republic (virtual meeting)
}

\begin{document}
\maketitle

%----------------------------
\section{Introduction}
%----------------------------

The $\nu_{\tau}$ and $\overline{\nu}_{\tau}$ particles were ones of last ingredients
of the Standard Model discovered experimentally \cite{DONUT1}.
So far only a few $\nu_\tau/\overline{\nu}_{\tau}$ neutrinos/antineutrinos were observed
experimentally. Recently, it was roughly estimated that about $300-1000$ neutrinos ($\nu_{\tau}+\overline{\nu}_{\tau}$)
will be observed by the SHiP (Search for Hidden Particles) experiment \cite{SHiP3,BR2018}. If so then it could 
considerably improve our knowledge in this weakly tested corner of the Standard Model.

The $\nu_\tau/\overline{\nu}_{\tau}$ neutrinos/antineutrinos are known to be primarily produced
from $D_s^{\pm}$ decays. The $D_s$ mesons are abundantly produced in proton-proton collisions.
Here we wish to make as realistic as possible predictions of the cross section
for production of $\nu_\tau/\overline{\nu}_{\tau}$ neutrinos/antineutrinos.
In our model $D_s^{\pm}$ mesons can be produced from both, charm and strange quark/antiquark
fragmentation, with a similar probability of the transition. The $s \to
D_s$ mechanism is expected to be especially
important at large rapidities (or large Feynman $x_F$) \cite{Goncalves:2018zzf}.  
%The $\tau$ leptons from the decay of $D_s$ mesons are polarized
%with respect to their direction of motion.
Here we wish to answer whether it has consequences
for forward production of neutrinos/antineutrinos for the SHiP experiment or not.

%----------------------------------------------------
\section{Some details of the approach}
%----------------------------------------------------

In our model we include two mechanisms of $D_s$ meson
production: $c \to D_s^+$, $\bar c \to D_s^-$, called leading fragmentation,
and $\bar s \to D_s^+$, $s \to D_s^-$, called subleading fragmentation.

The $c$ and $\bar c$ cross sections are calculated in the collinear NLO 
approximation using the \textsc{Fonll} framework \cite{FONLL}
or in the $k_t$-factorization approach \cite{kTfactorization}. Here, both 
the $gg$-fusion and $q\bar q$-annihilation production mechanisms for 
$c\bar c$-pairs with off-shell initial state partons are taken into 
consideration.

Not all charm hadrons must be created from the $c /{\bar c}$ fragmentation.
An extra hidden associated production of $c$ and $\bar c$
can occur in a complicated hadronization process. 
In principle, $c$ and $\bar c$ partons
can also hadronize into light mesons (e.g. kaons) with 
non-negligible fragmentation fraction (see e.g. Ref.~\cite{Epele:2018ewr}).
Similarly, fragmentation of light partons into heavy mesons may be well 
possible \cite{Kneesch:2007ey}.
In the present study we will discuss also results of 
\textsc{Pythia} hadronization 
to $D_s$ mesons in this context as well as our simple model of 
subleading fragmentation $s \to D_s^-$ and $\bar s \to D_s^+$ \cite{Goncalves:2018zzf}. 

The $s$ and $\bar s$ distributions are calculated here in the
  leading-order 
(LO) collinear factorization approach with on-shell initial state partons and with a special treatment of minijets at low transverse momenta, as adopted \textit{e.g.} in \textsc{Pythia}, 
by multiplyingcross section by a
suppression factor \cite{Sjostrand:2014zea}
\begin{equation}
F_{sup}(p_t) = \frac{p_t^4}{((p_{t}^{0})^{2} + p_t^2)^2}  \; .
\end{equation}
Within this framework the cross section of course strongly depends on the free parameter $p_{t}^{0}$ which could be, in principle, fitted to low energy charm experimental data \cite{Maciula:2017wov}. Here, we use rather conservative value $p_{t}^{0} = 1.5$ GeV. We use three different sets of the collinear parton distribution functions (PDFs): the MMHT2014 \cite{Harland-Lang:2014zoa} and the NNPDF30 \cite{Ball:2014uwa} parametrizations. Both of them provide an asymmetric strange sea quark distributions in the proton with $s(x)\neq \bar{s}(x)$. The dominant partonic mechanisms are $gs \to gs$, $g\bar s \to g\bar s$ (and their symmetric counterparts) and $gg \to s\bar s$.
In some numerical calculations we take into account also other $2\to 2$ diagrams with $s(\bar s)$-quarks in the final state, however, their contributions are found to be almost negligible.

The transition from quarks to hadrons in our calculations is done within the independent parton fragmentation picture.
Here, we follow the assumptions relevant for the case of low c.m.s. collision energies and/or small transverse momenta of hadrons,
as discussed in our recent analysis \cite{Maciula:2019iak}, and we assume that the hadron $H$ is emitted in the direction of parent
quark/antiquark $q$, i.e. $\eta_H = \eta_q$ (the same pseudorapidities or
polar angles). Within this approach we set the light-cone $z$-scaling, \textit{i.e.} we define $p_H^+ = z p_q^+$, where $p^{+} = E+p$.
In the numerical calculations we also include ``energy conservation'' conditions: $E_H > m_H$ and $E_{H} \leq E_{q}$.
If we take the parton as the only reservoir of energy (independent parton fragmentation) these conditions (especially the latter one)
may be strongly broken in the standard fragmentation framework with
constant rapidity $y_q = y_H$ scenario, especially, when discussing
small transverse momenta of hadrons. The light-cone scaling prescription reproduces the standard approach in the limit: $m_{q}, m_{H} \to 0$.

For $c/\bar c \to D_s^{\pm}$ fragmentation we take the traditional Peterson fragmentation function with $\varepsilon$ = 0.05.
In contrast to the standard mechanism, the fragmentation function for
$s/{\bar s} \to D_s^{\mp}$ transition is completely unknown which makes
the situation more difficult. For the case of light-to-light (light parton to light meson) transition rather softer fragmentation functions
(peaked at smaller $z$-values) are supported by phenomenological studies
\cite{Bertone:2017tyb}.
However, the light-to-heavy fragmentation should not be significantly different than for the heavy-to-heavy case.
The shape of the fragmentation function depends on mass of the hadron rather than on the mass of parton (see \textit{e.g.} Ref.~\cite{Kneesch:2007ey}). 
Therefore, here we take the same fragmentation function for the $s/{\bar s} \to D_s^{\mp}$ as for the
$c/\bar c \to D_s^{\pm}$. Besides the shape of the $s/{\bar s} \to D_s^{\mp}$ fragmentation function the relevant fragmentation fraction
is also unknown. The transition probability $P = P_{s \to D_s}$ can be treated as a free parameter and needs to be extracted
from experimental data. First attempt was done very recently in Ref.~\cite{Goncalves:2018zzf}, where $D^{+}_{s}/D^{-}_{s}$ production asymmetry was studied.

%-----------------------------------------------------------------------------
\begin{figure}[!htbp]
\begin{minipage}{0.47\textwidth}
 \centerline{\includegraphics[width=1.0\textwidth]{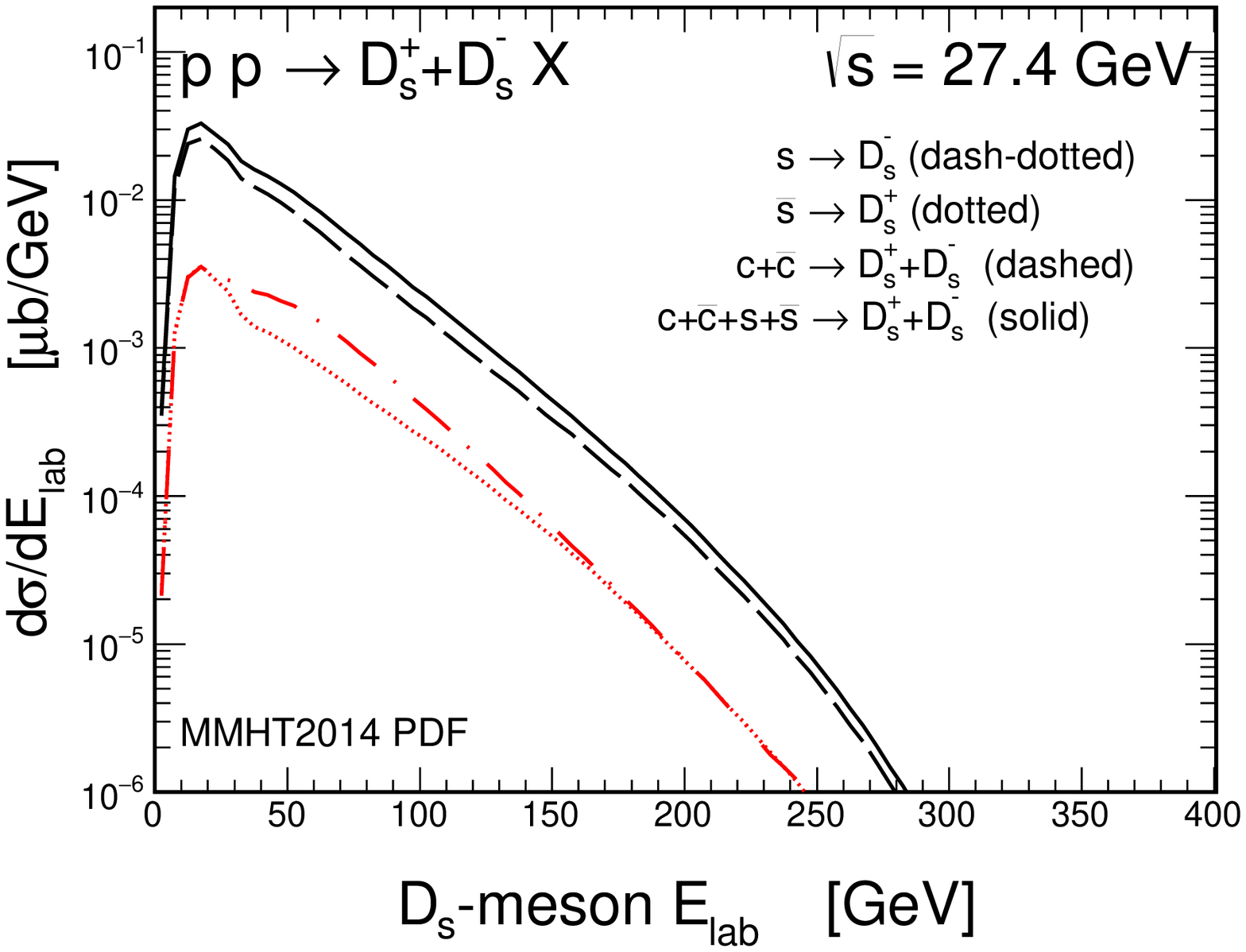}}
\end{minipage}
\hspace{0.5cm}
\begin{minipage}{0.47\textwidth}
 \centerline{\includegraphics[width=1.0\textwidth]{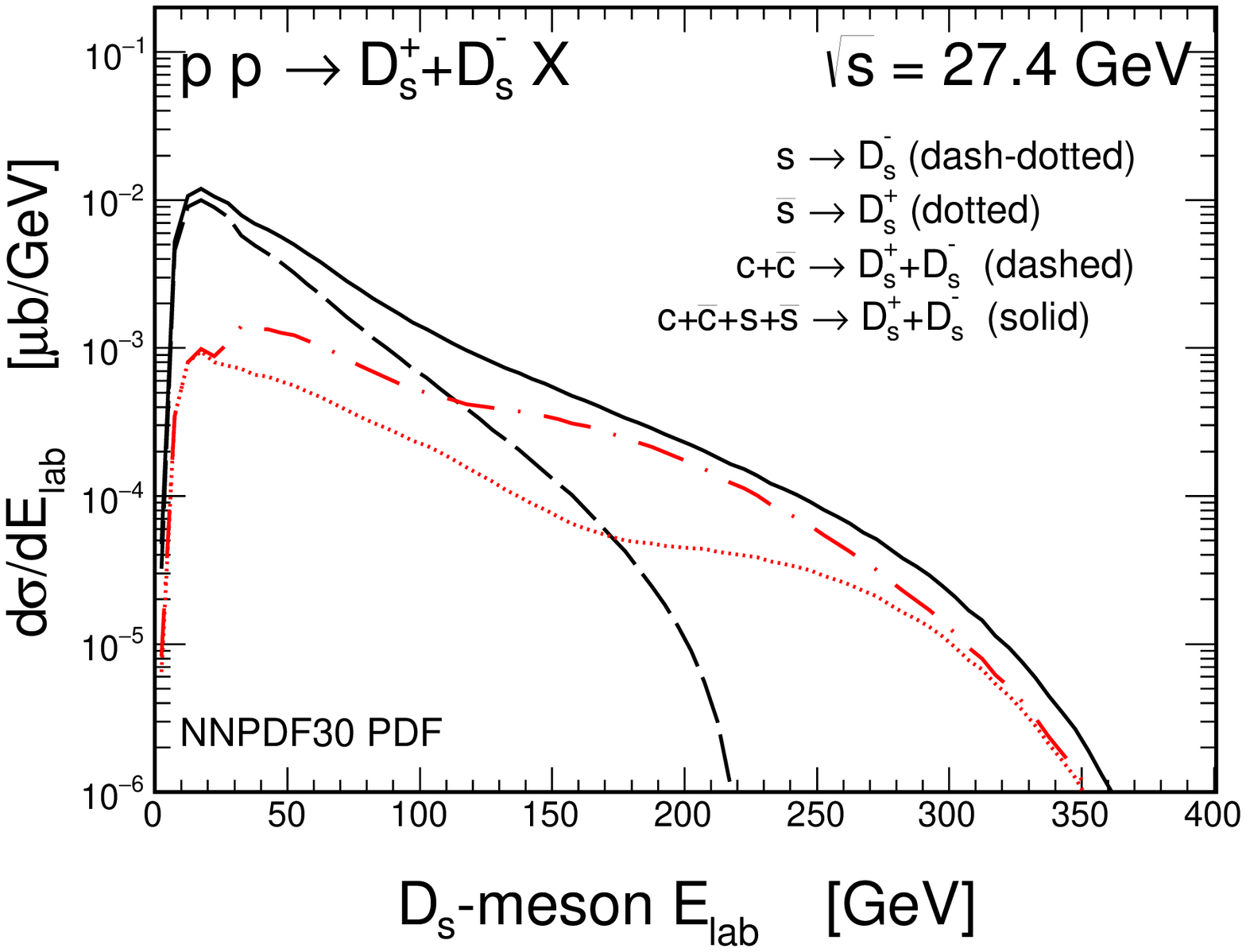}}
\end{minipage}
\caption{
\small Energy distributions of $D_{s}$ mesons in the laboratory frame for the MMHT2014 (left) and the NNPDF30 (right) collinear PDFs. Contributions from $c$ and $s$ quark fragmentation are shown separately.
}
 \label{fig:mesons_Elab}
\end{figure}
%------------------------------------------------------------------------------

In Fig.~\ref{fig:mesons_Elab} we show the energy distribution of $D_{s}$ mesons in the laboratory frame from proton-proton scattering at $\sqrt{s}=27.4$ GeV. Here we show separately the leading $c + \bar c \to D_{s}^{+} + D_{s}^{-}$ (dashed lines) and two subleading $s\to D_{s}^{-}$ (dash-dotted lines) and $s\to D_{s}^{-}$ (dotted lines) contributions as well as their sum $c + \bar c + s + \bar s \to D_{s}^{+} + D_{s}^{-}$ (solid lines). The left and right panels correspond to the MMHT2014 and the NNPDF30 PDFs, respectively.
In this calculation $P_{c \to D_s}=0.08$ and $P_{s \to D_s}=0.05$ were used.
A pretty much different results are obtained for the two different PDF sets, especially for large meson energies. Depending on the collinear PDFs used
our model leads to a rather small (the MMHT2014 PDF) or a
fairly significant (the NNPDF30 PDF) contribution to the $D_{s}$ meson production at large energies which comes from the $s/\bar s$-quark fragmentation. A future measurement of $D_{s}$ mesons at low energies would help to better understand underlying mechanism and improve predictions for $\nu_{\tau}/\overline{\nu}_{\tau}$ production for the SHiP experiment.

The considered here decay channels: $D_s^+ \to \tau^+ \nu_{\tau}$ and $D_s^- \to \tau^- {\overline \nu}_{\tau}$, which are the sources of the direct neutrinos,
are analogous to the standard text book cases of $\pi^+ \to \mu^+
\nu_{\mu}$ and $\pi^- \to \mu^- {\overline \nu}_{\mu}$ decays, discussed in detail in the past (see e.g. Ref~\cite{Renton}). The same formalism used for the pion decay applies also to the $D_s$ meson decays.
Since pion has spin zero it decays isotropically in its rest frame. However, the produced muons are polarized in its direction of motion
which is due to the structure of weak interaction in the Standard Model. The same is true for $D_s^{\pm}$ decays and polarization of
$\tau^{\pm}$ leptons.To calculate cross section for $\nu_{\tau}/{\overline \nu}_{\tau}$ production
the decay branching fraction BR$(D_s^{\pm} \to \tau^{\pm} \nu_{\tau}/{\overline \nu}_{\tau})$ = 0.0548$\pm$
0.0023 \cite{PDG} must be included.

The $\tau$ decays are rather complicated due to having many possible decay 
channels~\cite{PDG}. Nevertheless, all confirmed decays lead to production of 
$\nu_{\tau}$ (${\overline \nu}_{\tau}$). This means total amount of neutrinos/antineutrinos produced from $D_s$ decays into $\tau$ lepton 
is equal to the amount of antineutrinos/neutrinos produced in subsequent $\tau$ decay.
But, their energy distributions will be different due to $D_s$ 
production asymmetry in the case of the subleading fragmentation mechanism.

The purely leptonic channels, analogous to the
$\mu^{\pm} \to e^{\pm} ({\overline \nu}_{\mu}/\nu_{\mu})( \nu_e / {\overline \nu}_e)$
decay cover only about 35\% of all 
$\tau$ lepton decays. Remaining 65\% are semi-leptonic decays. 
They differ quite drastically from each other and each gives 
slightly different energy distribution for $\nu_{\tau}$ (${\overline \nu}_{\tau}$).
In our model for the decay of $D_s$ mesons there
is almost full polarization of $\tau$ particles with respect to the direction of their motion.
The mass of the $\tau$ lepton ($1.777$ GeV) is very similar as the mass of the $D_s$
meson ($1.968$ GeV). Therefore, direct neutrino takes away only a small fraction of
energy/momentum of the mother $D_s$. In this calculation we use \textsc{Tauola} code \cite{TAUOLA}.

In the case of the SHiP experiment a dedicated lead target was proposed.
At not too small energies ($\sqrt{s_{NN}} > 5$ GeV), the cross section for 
$\nu_{\tau} Pb$ and $\overline{\nu}_{\tau} Pb$ interactions
can be obtained from elementary cross sections as:
$\sigma(\nu_{\tau} Pb) = Z \sigma(\nu_{\tau} p) + (A-Z) 
\sigma(\nu_{\tau} n)$, and
$\sigma(\overline{\nu}_{\tau} Pb) =
Z \sigma(\overline{\nu}_{\tau} p) + (A-Z) \sigma(\overline{\nu}_{\tau} n)$.
Shadowing effects depend on $x$ variable 
(parton longitudinal momentum fraction),
i.e. on neutrino/antineutrino energy.
At not too high energies (not too small $x$) shadowing effects 
are rather small and can be neglected at present accuracy
having in mind other uncertainties. On the other hand for the $x$-ranges considered here
the antishadowing and/or EMC-effect may appear non-negligible but still rather small and shall not affect the numerical predictions presented here.

%Both elementary as well as nuclear cross sections strongly depend on
%neutrino/antineutrino energy \cite{JR2010}. For $\tau$
%neutrino/antineutrino interactions there is also an energy threshold
%related to the mass of $\tau^{\pm}$ 
%which reduces cross section compared to
%$\nu_{\mu}/{\overline{\nu}_{\mu}}$ 
%and practically cuts off contributions of nucleon resonances. Therefore 
%one should include practically only deep-inelastic region.

The probability of interacting of neutrino with the lead
target can be calculated as:
\begin{equation}
P_{\nu_{\tau}/\overline{\nu}_{\tau}}^{\mathrm{target}}(E) = \int_0^{d} n_{\mathrm{cen}} \sigma_{\nu_{\tau} Pb}(E) dz 
= n_{\mathrm{cen}} \sigma_{\nu_{\tau} Pb}(E) d
\; ,
\label{probability_of_interaction}
\end{equation}
where $n_{\mathrm{cen}}$ is a number of scattering centers (lead nuclei) per volume 
element and the target thickness is $d \approx$ 2 m \cite{SHiP3}.
Using the \textsc{NuWro} Monte Carlo generator \cite{NuWro}, we obtain   
$\sigma(E)/E \sim 1.09 \times 10^{-38}$ cm$^2/$GeV for neutrino and
$0.41 \times 10^{-38}$ cm$^2/$GeV for antineutrino
for the $E = 100$ GeV. The number of scattering centers is
$n_{\mathrm{cen}} = (11.340 / 207.2) N_A$, where $N_A$ = 6.02 $\times$ 10$^{23}$ is the Avogadro number.

The energy dependent flux of neutrinos can be written as:
\begin{equation}
\Phi_{\nu_{\tau}/\overline{\nu}_{\tau}}(E) =
\frac{N_p}{\sigma_{pA}} d\sigma_{pA \to \nu_{\tau}}(E)/dE \; ,
\label{neutrino_flux}
\end{equation}
where $N_p$ is integrated number of beam protons ($N_p = 2 \times 10^{20}$
according to the current SHiP project).
The $\sigma_{pA}$ in Eq.~(\ref{neutrino_flux}) is a crucial quantity which 
requires a short disscusion.
In Ref.~\cite{BR2018} it was taken as
$\sigma_{pA} = A \cdot \sigma_{pN}$ where $\sigma_{pN}$ = 10.7 was used. 
We do not know the origin of this number.
Naively $\sigma_{pN}$ should be the inelastic $pN$ cross section.

Finally the number of $\nu_{\tau}$ or $\overline{\nu}_{\tau}$ observed
in the $Pb$ target is calculated from the formula:
\begin{equation}
N_{\nu_{\tau}/\overline{\nu}_{\tau}}^{\mathrm{target}} =
\int dE \Phi_{\nu_{\tau}/\overline{\nu}_{\tau}}(E)
P_{\nu_{\tau}/\overline{\nu}_{\tau}}^{\mathrm{target}}(E) \; .
\label{number_of_observed_neutrinos}
\end{equation}
Here $\Phi_{\nu_{\tau}/\overline{\nu}_{\tau}}(E)$ is
calculated from different approaches to $D_s$ meson production including their subsequent decays
and $P_{\nu_{\tau}/\overline{\nu}_{\tau}}^{\mathrm{target}}(E)$ is obtained
using Eq.(\ref{probability_of_interaction}). 

\section{Numerical results}

\begin{figure}[!h]
\begin{minipage}{0.47\textwidth}
  \centerline{\includegraphics[width=1.0\textwidth]{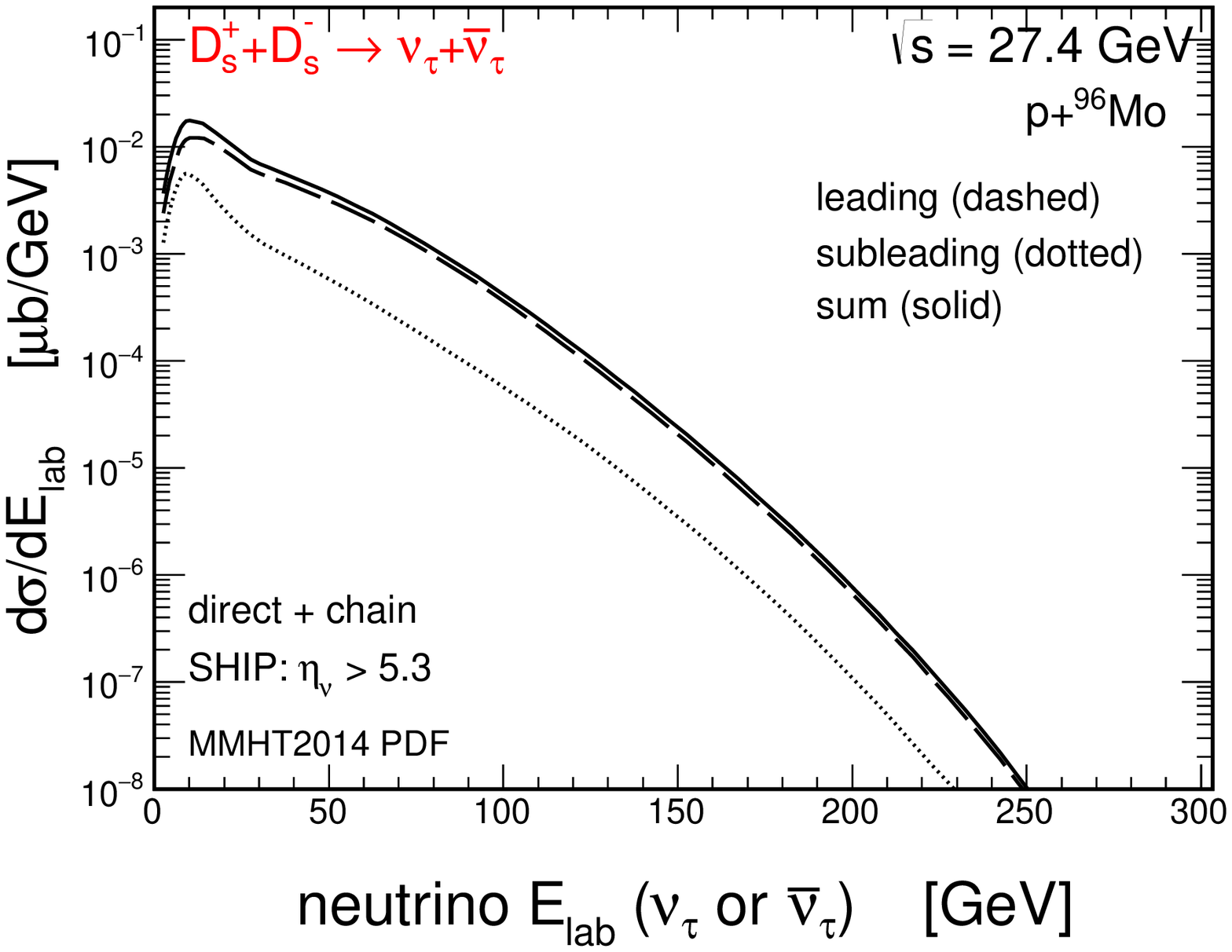}}
\end{minipage}
\begin{minipage}{0.47\textwidth}
  \centerline{\includegraphics[width=1.0\textwidth]{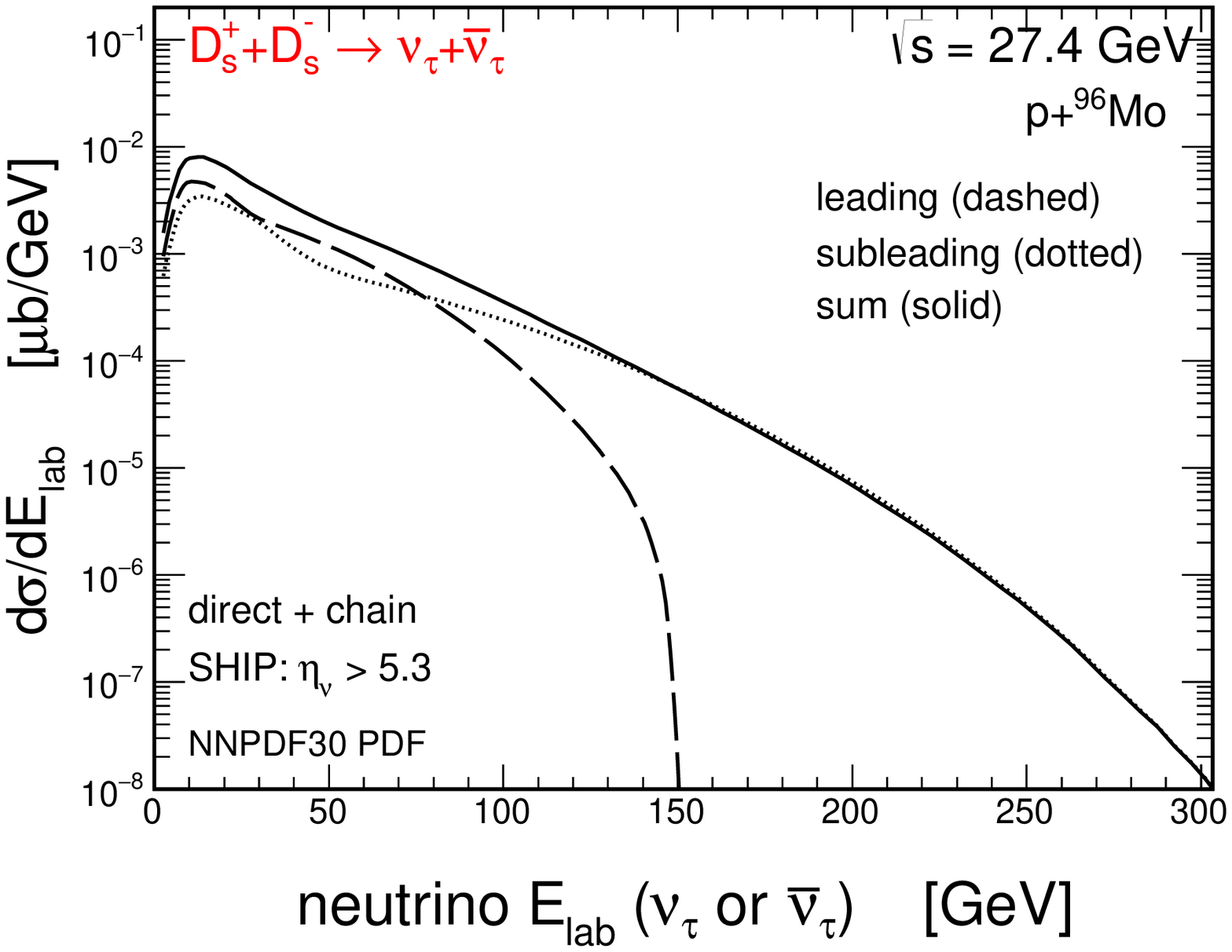}}
\end{minipage}
  \caption{
\small Laboratory frame energy distributions of $\nu_{\tau}$ (or $\overline{\nu}_{\tau}$) neutrinos for MMHT2014 (left) and NNPDF30 (right) sets of collinear PDFs.
Here we show in the same panel the leading and subleading contributions
as well as their sum.
}
\label{fig:nu_Elab_sum}
\end{figure}

In Fig.~\ref{fig:nu_Elab_sum} we show the impact of the subleading contribution
for the predictions of $\nu_{\tau}$ and/or $\overline{\nu}_{\tau}$ energy distributions for the SHiP experiment. Again we obtain two different scenarios for the two different PDF sets.
The MMHT2014 PDFs set leads to an almost negligible subleading contribution in the whole energy range while the NNPDF30 PDFs set provides the subleading contribution to be dominant
at larger energies ($E_{\mathrm{lab}} > 100$ GeV). If such distributions could be measured by the SHiP then 
they could be useful to constrain the PDFs in the purely known kinematical region.

%----------------------------------------------------------------------------
%----------------------------------------------------------------------------

Predictions for observed numbers of neutrinos/antineutrinos for the SHIP experiment are collected in Table \ref{tab:number_nu}.
Quite different numbers are obtained for the different considered scenarios.
We have predicted $\sim 800 - 2000$ tau neutrino events from charm quark fragmentation 
and $\sim 200-400$ tau neutrino events from strange quark fragmentation. The subleading fragmentation may increase
the probability of observing $\nu_{\tau}/\overline{\nu}_{\tau}$ neutrinos/antineutrinos.
We get larger numbers than in Ref.~\cite{BR2018} but smaller than in Ref.~\cite{SHiP3}.
The chain contribution is significantly larger (about factor 7) than the direct one.
For the MMHT2014 distribution the contribution of the leading mechanism is much larger than
for the subleading one. For the NNPDF30 distributions the situation is reversed.
We predict large observation asymmetry (see the last column) for 
$\nu_{\tau}$ and $\overline{\nu}_{\tau}$. This asymmetry is bigger than
shown e.g. in Refs.~\cite{BR2018,SHiP3}. This is due to the subleading mechanism for $D_{s}^{\pm}$ meson production included in the present paper.
The observation asymmetry for the leading contribution which comes from the differences of the $\nu_{\tau}$ and $\overline{\nu}_{\tau}$ interactions with target are estimated at the level of 50\%. In the case of the subleading contribution the asymmetry increases to 60-70\%.
More details of the study can be found in original article \cite{Maciula:2019clb}.

%------------------------------------------------------------------------------------------------------------------------------
\begin{table}[tb]%
\caption{Number of observed $\nu_{\tau}$ and $\overline{\nu}_{\tau}$ for the SHiP experiment.}

\label{tab:number_nu}
\centering %
%\newcolumntype{Z}{>{\centering\arraybackslash}X}
%\newcommand{\tn}{\tabularnewline}
%\resizebox{\textwidth}{!}{%
\begin{tabularx}{1.0\linewidth}{c c c c c c}
\\[-3.5ex] 
\toprule[0.1em] %
%\\[-4.ex] 
%\\[1.0ex]

\multirow{2}{6.cm}{Framework/mechanism} & \multicolumn{5}{c}{\multirow{1}{6.cm}{Number of observed neutrinos}}  \\ [-0.ex]
\multirow{2}{6.cm}{}   & \multirow{1}{1.5cm}{flavour}  & \multirow{1}{1.5cm}{direct} & \multirow{1}{1.5cm}{chain} & \multirow{1}{1.5cm}{$\nu_{\tau} + \overline{\nu}_{\tau}$} & \multirow{1}{1.5cm}{$\frac{\nu_{\tau} - \overline{\nu}_{\tau}}{\nu_{\tau} + \overline{\nu}_{\tau}}$} \\ [+0.1ex]

\bottomrule[0.1em]

   \multirow{1}{6.cm}{FONLL + NNPDF30 NLO PDF}      &   \multirow{1}{1.5cm}{$\nu_{\tau}$}  &   \multirow{1}{1.5cm}{96}  &  \multirow{1}{1.5cm}{515}  & \multirow{2}{1.5cm}{818} & \multirow{2}{1.5cm}{0.49} \\[-1.0ex]
 \multirow{1}{6.cm}{$c/\bar{c} \to D_{s}^{\pm} \to \nu_{\tau}/\overline{\nu}_{\tau}$}      &   \multirow{1}{1.5cm}{$\overline{\nu}_{\tau}$}  &   \multirow{1}{1.5cm}{27}  &  \multirow{1}{1.5cm}{180}  & \multirow{2}{1.5cm}{} & \multirow{2}{1.5cm}{}  \\
 \hline
   \multirow{1}{6.cm}{LO coll. + NNPDF30 LO PDF}      &   \multirow{1}{1.5cm}{$\nu_{\tau}$}  &   \multirow{1}{1.5cm}{93}  &  \multirow{1}{1.5cm}{1092}  & \multirow{2}{1.5cm}{1416} & \multirow{2}{1.5cm}{0.67}   \\[-1.0ex]
 \multirow{1}{6.cm}{$s/\bar{s} \to D_{s}^{\pm} \to \nu_{\tau}/\overline{\nu}_{\tau}$}      &   \multirow{1}{1.5cm}{$\overline{\nu}_{\tau}$}  &   \multirow{1}{1.5cm}{75}  &  \multirow{1}{1.5cm}{156} & \multirow{2}{1.5cm}{}     \\
\bottomrule[0.1em]
 \multirow{1}{6.cm}{FONLL + MMHT2014nlo PDF}      &   \multirow{1}{1.5cm}{$\nu_{\tau}$}  &   \multirow{1}{1.5cm}{277}  &  \multirow{1}{1.5cm}{1427} & \multirow{2}{1.5cm}{2292}       & \multirow{2}{1.5cm}{0.49} \\[-1.0ex]
 \multirow{1}{6.cm}{$c/\bar{c} \to D_{s}^{\pm} \to \nu_{\tau}/\overline{\nu}_{\tau}$}      &   \multirow{1}{1.5cm}{$\overline{\nu}_{\tau}$}  &   \multirow{1}{1.5cm}{80}  &  \multirow{1}{1.5cm}{508} & \multirow{2}{1.5cm}{}  & \multirow{2}{1.5cm}{}    \\
 \hline
  \multirow{1}{6.cm}{LO coll. + MMHT2014lo PDF}      &   \multirow{1}{1.5cm}{$\nu_{\tau}$}  &   \multirow{1}{1.5cm}{59}  &  \multirow{1}{1.5cm}{435}  & \multirow{2}{1.5cm}{632}     & \multirow{2}{1.5cm}{0.56} \\[-1.0ex]
 \multirow{1}{6.cm}{$s/\bar{s} \to D_{s}^{\pm} \to \nu_{\tau}/\overline{\nu}_{\tau}$}      &   \multirow{1}{1.5cm}{$\overline{\nu}_{\tau}$}  &   \multirow{1}{1.5cm}{21}  &  \multirow{1}{1.5cm}{117} & \multirow{2}{1.5cm}{} & \multirow{2}{1.5cm}{}     \\
%\bottomrule[0.1em]
%  \multirow{1}{7.cm}{FONLL + MMHT2014nlo PDF}      &   \multirow{1}{1.5cm}{$\nu_{\tau}$}  &   \multirow{1}{1.5cm}{277}  &  \multirow{1}{1.5cm}{2139} & \multirow{2}{1.5cm}{3259} & \multirow{2}{1.5cm}{0.48}      \\[-1.0ex]
% \multirow{1}{7.cm}{$c/\bar{c} \to D_{s}^{\pm} \to \nu_{\tau}/\overline{\nu}_{\tau}$ (polarization off)}      &   \multirow{1}{1.5cm}{$\overline{\nu}_{\tau}$}  &   \multirow{1}{1.5cm}{80}  &  \multirow{1}{1.5cm}{763}  & \multirow{2}{1.5cm}{}  & \multirow{2}{1.5cm}{}  \\
% \hline
%  \multirow{1}{7.cm}{LO coll. + MMHT2014lo PDF}      &   \multirow{1}{1.5cm}{$\nu_{\tau}$}  &   \multirow{1}{1.5cm}{60}  &  \multirow{1}{1.5cm}{562}   & \multirow{2}{1.5cm}{844}  & \multirow{2}{1.5cm}{0.47} \\[-1.0ex]
% \multirow{1}{7.cm}{$s/\bar{s} \to D_{s}^{\pm} \to \nu_{\tau}/\overline{\nu}_{\tau}$ (polarization off)}      &   \multirow{1}{1.5cm}{$\overline{\nu}_{\tau}$}  &   \multirow{1}{1.5cm}{22}  &  \multirow{1}{1.5cm}{200} & \multirow{2}{1.5cm}{} & \multirow{2}{1.5cm}{}      \\

\bottomrule[0.1em]
\end{tabularx}
%}
\end{table}
%-------------------------------------------------------------------------------------------------------------------------------

\end{document}